\documentclass[amsmath,amssymb,prb]{revtex4-2}

\usepackage{graphicx}
\usepackage{dcolumn}
\usepackage{bm}

\begin{document}

\title{Stress calculation in linear scaling DFT: convergence and dynamics}

\author{Shereif Y. Mujahed}
\affiliation{Dept. Physics \& Astronomy, University College London, Gower Street, London, WC1E 6BT, United Kingdom.}
\affiliation{London Centre for Nanotechnology, University College London, 17-19 Gordon Street, London, WC1H 0AH, United Kingdom.}

\author{Tsuyoshi Miyazaki}
\affiliation{Research Centre for Materials Nanoarchitectonics (WPI-MANA), National Institute for Materials Science (NIMS), 1-1 Namiki, Tsukuba, Ibaraki 305-0044, Japan}

\author{David R. Bowler}%
\email{david.bowler@ucl.ac.uk}
\affiliation{London Centre for Nanotechnology, University College London, 17-19 Gordon Street, London, WC1H 0AH, United Kingdom.}
\affiliation{Research Centre for Materials Nanoarchitectonics (WPI-MANA), National Institute for Materials Science (NIMS), 1-1 Namiki, Tsukuba, Ibaraki 305-0044, Japan}
\affiliation{Dept. Physics \& Astronomy, University College London, Gower Street, London, WC1E 6BT, United Kingdom.}

\date{\today}

\begin{abstract}
  We present the approach needed to calculate stress within density functional theory (DFT) using a localised orbital basis, both for exact diagonalisation and linear scaling approaches, and demonstrate our implementation within the large scale DFT code \textsc{Conquest}.  For the linear scaling approach, we test the rate of convergence of stress with density matrix range, and compare it to the convergence of energy and forces for different materials with a range of band gaps.  We show that excellent convergence is found for modest cutoffs (below 0.1GPa error for range of 20a$_{0}$), and show that large-scale isothermal-isobaric molecular dynamics is stable and accurate, showing negligible drift over picoseconds of dynamics.
\end{abstract}

\maketitle

\section{Introduction}

Density functional theory (DFT) has become the method of choice for electronic structure calculations\cite{Martin:2020lk}.  In condensed matter research, plane wave basis sets have often been the basis of choice, which has tended to limit the size of all but the largest simulations to hundreds of atoms\cite{Carnimeo:2023ko}; the adoption of local orbital basis sets has enabled larger scale simulations, both with exact diagonalisation and with linear scaling methods to find the ground state\cite{Ratcliff:2020gd,Garcia:2020lt,Prentice:2020yb,Kuhne:2020pc,Nakata:2020dn}.  The \textsc{Conquest} code\cite{Nakata:2020dn} has demonstrated simulations of several thousand atoms with exact diagonalisation\cite{Nakata:2022lb}, and millions of atoms with linear scaling\cite{Bowler:2010uq,Arita:2014qr}.

Simulations of condensed and soft matter systems require that the forces on the ions can be efficiently calculated as exact differentials of the energy; in many cases, the stresses on the simulation cell are also required.  We have discussed the calculation of forces in \textsc{Conquest} for both exact diagonalisation and linear scaling before\cite{Miyazaki:2004ee}, and noted that the motion of basis functions with the ions requires evaluation of Pulay terms\cite{Pulay:1969le} which are readily found analytically.  The evaluation of stresses is standard in plane-wave DFT\cite{Martin:2020lk} though for plane-wave basis set there is a missing term which arises from the change of basis with simulation cell size (a Pulay stress), and cannot be calculated analytically, leading to a need to increase basis set size for stress evaluation.  The calculation of stress for local orbital basis sets has been shown before\cite{Feibelman:1991we,Soler:2002kn} but the calculation of stress with linear scaling, and the convergence of the stress, has only been investigated in the context of finite electronic temperature simulations\cite{Sharma:2020no}.  We reported preliminary results\cite{Nakata:2020dn} on stress with linear scaling methods, but gave only an overview of the approach; below, we give full details, and note an important correction to our preliminary results.  We also demonstrate stable NPT molecular dynamics.

The paper is organised as follows.  We recap the expressions required to evaluate the energy both in exact diagonalisation and linear scaling modes; we note an important constraint that is applied to maintain electron number in linear scaling calculations.  We then examine the approach needed to calculate stresses with localised orbitals and the linear scaling approach in \textsc{Conquest}.  We conclude by presenting various tests of the implementation; first, testing the convergence of energy, forces and stresses with density matrix range, and then demonstrating stable molecular dynamics in different ensembles.

\section{Energies}
\label{sec:energies}

For a system with total energy $E$ we define the forces on the atoms and the stresses on the simulation cell in the usual way, with $\mathbf{F}_{i} = -\nabla_{i} E$ for atom $i$, and $\sigma_{\lambda\mu} = (1/\Omega)\partial E/\partial \epsilon_{\lambda\mu}$, for the strain $\epsilon_{\lambda\mu}$ (with $\lambda$ and $\mu$ each representing one of $x$, $y$ or $z$) and $\Omega$ the simulation cell volume.  This definition of stress has units of pressure; an alternative definition, often known as the internal stress, omits the prefactor of $1/\Omega$  to give units of energy\cite{Nielsen:1983xu,Nielsen:1985cz}.

In order to implement accurate forces and stresses, we have to consider the details of how energies are calculated, and ensure that the forces and stresses are exact derivatives of the energy.
In \textsc{Conquest}, in common with other local orbital pseudopotential codes such as Siesta\cite{Soler:2002kn} and OpenMX\cite{Ozaki:2005dp}, we use a reformulation of the electrostatic part of the total energy based on neutral atom charges.  By associating a pseudo-atomic density $n^{(a)}_{i}(\mathbf{r})$ with each atom\footnote{We have a pseudo-atomic density, since we use a pseudopotential and solve for pseudo-atomic orbitals (PAOs).}, we can rewrite the Hartree energy of the electrons and the ionic core-core energy, avoiding any long range terms, leaving only a screened core-core interation (SCC).  We combine the local part of the pseudopotential and the Hartree potential from the pseudo-atomic density to give a neutral atom potential: $V_{NA}(\mathbf{r}) = \sum_{i} V_{PS, Loc, i}(\mathbf{r}) + V^{(a)}_{Ha,i}(\mathbf{r})$.  We define the total pseudo-atomic density (PAD) for the system as $n^{PAD}(\mathbf{r}) = \sum_{i} n^{(a)}_{i}(\mathbf{r})$, and a charge density difference between the ground state charge density and the PAD as $\delta n(\mathbf{r}) = n(\mathbf{r}) - n^{PAD}(\mathbf{r})$.  This density difference gives a potential $V_{\delta Ha}$ and an energy $E_{\delta Ha}$, and we define the Harris-Foulkes\cite{Harris1985,Foulkes1989} and Kohn-Sham\cite{Kohn:1965bq} energies below in Eq.~\eqref{eq:4} and Eq.~\eqref{eq:34}, with a Hamiltonian written as:
\begin{eqnarray}
  \label{eq:1}
  \hat{H} &=& \left(-\frac{1}{2}\nabla^{2} + \hat{V}_{NA} + \hat{V}_{PS, NL} + V_{\delta Ha} + V_{XC}\right)
\end{eqnarray}
where $\hat{V}_{PS, NL}$ is the non-local part of the pseudopotential.

\textsc{Conquest} uses a basis set of localised functions, known as support functions\cite{Hernandez:1996bf}, to represent the Kohn-Sham eigenstates.  Most commonly these functions are represented by pseudo-atomic orbitals (PAOs) which are generated by solving the Schr\"{o}dinger equation for isolated atoms, using pseudopotentials\cite{Bowler:2019fv}.  The eigenstates are then expressed as: $\vert\psi_{n}\rangle = \sum_{i\alpha} c^{i\alpha}_{n}\vert\phi_{i\alpha}\rangle,$
where $i\alpha$ indicates the support functions on atom $i$, with the support functions indexed by $\alpha$.  Using these basis functions, the Schr\"{o}dinger equation can be re-written as a matrix equation, $H_{i\alpha j\beta}c^{j\beta}_{n} = \epsilon_{n}S_{i\alpha j\beta}c^{j\beta}_{n}$, noting that the support functions are non-orthogonal, with overlap matrix $S_{i\alpha j\beta}$, and the matrices defined in the usual way: $H_{i\alpha j\beta} = \langle \phi_{i\alpha}\vert\hat{H}\vert\phi_{j\beta}\rangle$ and $S_{i\alpha,j\beta} = \langle\phi_{i\alpha}\vert\phi_{j\beta}\rangle$.  The individual terms in the energy can be defined either in terms of the eigenvalues and eigenstates, or in terms of the Hamiltonian and density matrices; the latter approach, coupled with sparse matrices, enables a linear scaling formulation of DFT to be constructed\cite{Bowler:2012zt,Goedecker:1999pv}.

\subsection{Energy in terms of eigenvalues and eigenstates}
\label{sec:energy-terms-eigenv}

We present first the Harris-Foulkes (HF) and Kohn-Sham (KS) energies used when solving by diagonalisation; below we turn to the linear scaling formulation.  We also consider the inclusion of non-linear partial core corrections\cite{Louie:1982hh} which include an additional charge density in the core region of each atom, $n^{C}_{i}(\mathbf{r})$, in the calculation of the exchange-correlation energy, and for simplicity we define $\tilde{n}(\mathbf{r}) = n(\mathbf{r}) + n^{C}(\mathbf{r})$, where $n^{C}(\mathbf{r}) = \sum_{i} n^{C}_{i}(\mathbf{r})$. 
\begin{eqnarray}
  \label{eq:4}
  E_{HF} &=& E_{Band} + \Delta E_{Ha} + \Delta E_{XC} + E_{SCC}\\
  E_{HF} &=& \sum_{n} 2f_{n} \epsilon_{n}+\left[-\frac{1}{2}\int d\mathbf{r} \delta V_{Ha}\left(\mathbf{r}\right) \delta n(\mathbf{r}) - \int d\mathbf{r} \delta V_{Ha}\left(\mathbf{r}\right) n^{PAD}(\mathbf{r})\right] \nonumber\\
  &+& \int d\mathbf{r} \left(\tilde{n}(\mathbf{r})\epsilon_{XC}\left[\tilde{n}(\mathbf{r})\right] - n(\mathbf{r})\mu_{XC}\left[\tilde{n}(\mathbf{r})\right]\right)+
      \frac{1}{2}\left(\sum_{i,j}\frac{Z_{i}Z_{j}}{\left\vert\mathbf{R}_{i} - \mathbf{R}_{j}\right\vert} - \int d\mathbf{r} V^{(a)}_{Ha,j}(\mathbf{r})n^{(a)}_{i}\left(\mathbf{r}\right)\right)\\
  \label{eq:34}
  E_{KS} &=& E_{KE} + E_{NA} + E_{PS, NL} + E_{\delta Ha} + E_{XC} + E_{SCC}\\
  E_{KS} &=& -\frac{1}{2}\sum_{n} 2f_{n}\langle \psi_{n} \left\vert \nabla^{2}\right\vert \psi_{n}\rangle + \int d\mathbf{r} V_{NA}(\mathbf{r})n(\mathbf{r}) + \sum_{n} 2f_{n} \langle\psi_{n}\vert\hat{V}_{PS, NL}\vert\psi_{n}\rangle \nonumber\\
         &+& \frac{1}{2}\int d\mathbf{r} \delta n(\mathbf{r})\delta V_{Ha}(\mathbf{r}) + \int d\mathbf{r} \tilde{n}(\mathbf{r})\epsilon_{XC}\left[\tilde{n}(\mathbf{r})\right]+
             \frac{1}{2}\left(\sum_{i,j}\frac{Z_{i}Z_{j}}{\left\vert\mathbf{R}_{i} - \mathbf{R}_{j}\right\vert} - \int d\mathbf{r} V^{(a)}_{Ha,j}(\mathbf{r})n^{(a)}_{i}\left(\mathbf{r}\right)\right)
\end{eqnarray}
where $\epsilon_{n}$ are the eigenvalues indexed with subscript $n$, $f_{n}$ are the occupancies of the eigenstates and the non-local pseudopotential is defined as $\hat{V}_{PS, NL} = \sum_{i} \sum_{\gamma}\vert\chi_{i\gamma}\rangle A_{\gamma} \langle \chi_{i\gamma}\vert$, summing over atoms $i$ and projectors $\gamma$ in the usual way\cite{Hamann:2013kq}.

When operating without self-consistency\cite{Torralba:2009nr}, we use the fixed superposition of atomic densities defined above as the density, so that $n(\mathbf{r}) = n^{PAD}(\mathbf{r})$, which fits well with the neutral-atom approach.  In this case, only the Harris-Foulkes energy is reliable, and $\Delta E_{Ha} = 0$ as $\delta n(\mathbf{r}) = 0$, while $\Delta E_{XC}$ is found as usual, but with $\tilde{n}(\mathbf{r}) = n^{PAD}(\mathbf{r}) + n^{C}(\mathbf{r})$.

\subsection{Energy in terms of the density matrix}
\label{sec:basis-sets-density}

For linear scaling operation, all calculations are performed in terms of the density matrix and the charge density\cite{Bowler:2002pt,Bowler:2010uq}.  
The density matrix in real space is defined in terms of the support functions and a matrix $K^{i\alpha j\beta}$ as $\rho(\mathbf{r},\mathbf{r}^{\prime}) = \sum_{i\alpha j\beta} \phi_{i\alpha}(\mathbf{r}) K^{i\alpha j\beta} \phi_{j\beta}(\mathbf{r}^{\prime})$, and the connection to eigenvector coefficients can be seen in the formal definition of $K^{i\alpha j\beta} = \sum_{n} f_{n}c^{i\alpha\star}_{n} c^{j\beta}_{n}$.  The charge density is then written as $n(\mathbf{r}) = \sum_{i\alpha j\beta} \phi_{i\alpha}(\mathbf{r}) K^{i\alpha j\beta} \phi_{j\beta}(\mathbf{r})$ and the energies are given by:
\begin{eqnarray}
  \label{eq:32}
  E_{HF} &=& \sum_{n} 2\mathrm{Tr}[KH]+\left[-\frac{1}{2}\int d\mathbf{r} \delta V_{Ha}\left(\mathbf{r}\right) \delta n(\mathbf{r}) - \int d\mathbf{r} \delta V_{Ha}\left(\mathbf{r}\right) n^{PAD}(\mathbf{r})\right] \nonumber\\
  &+& \int d\mathbf{r} \left(\tilde{n}(\mathbf{r})\epsilon_{XC}\left[\tilde{n}(\mathbf{r})\right] - n(\mathbf{r})\mu_{XC}\left[\tilde{n}(\mathbf{r})\right]\right)+
  \frac{1}{2}\left(\sum_{i,j}\frac{Z_{i}Z_{j}}{\left\vert\mathbf{R}_{i} - \mathbf{R}_{j}\right\vert} - \int d\mathbf{r} V^{(a)}_{Ha,j}(\mathbf{r})n^{(a)}_{i}\left(\mathbf{r}\right)\right)\\
  E_{KS} &=& -\frac{1}{2} \sum_{i\alpha j\beta}2K^{j\beta i\alpha}\langle \phi_{i\alpha} \left\vert \nabla^{2}\right\vert \phi_{j\beta}\rangle  + \int d\mathbf{r} V_{NA}(\mathbf{r})n(\mathbf{r}) + \sum_{i\alpha j\beta} 2K^{j\beta i\alpha}\langle\phi_{i\alpha}\vert\hat{V}_{PS, NL}\vert\phi_{j\beta}\rangle \nonumber\\
         &+& \frac{1}{2}\int d\mathbf{r} \delta n(\mathbf{r})\delta V_{Ha}(\mathbf{r}) + \int d\mathbf{r} \tilde{n}(\mathbf{r})\epsilon_{XC}\left[\tilde{n}(\mathbf{r})\right]+
             \frac{1}{2}\left(\sum_{i,j}\frac{Z_{i}Z_{j}}{\left\vert\mathbf{R}_{i} - \mathbf{R}_{j}\right\vert} - \int d\mathbf{r} V^{(a)}_{Ha,j}(\mathbf{r})n^{(a)}_{i}\left(\mathbf{r}\right)\right)  
\end{eqnarray}

For relatively small systems (up to a few thousand atoms) the ground state is found by direct diagonalisation of the Hamiltonian\cite{Nakata:2020dn}.  To go beyond systems of this size, linear scaling methods are needed\cite{Bowler:2012zt}, as described in the next section.

\subsection{Linear scaling and electron number}

The ground state density matrix of semiconducting and insulating systems (i.e. systems with a gap) can be found with computational effort that scales linearly with system size\cite{Bowler:2012zt,Goedecker:1999pv} by working with localised functions as basis functions, and restricting the range of the density matrix (a controlled approximation).  Within \textsc{Conquest}, a variational minimisation of the energy with respect to the density matrix elements is performed, while maintaining two properties of the system: idempotency of the density matrix; and correct total electron number.

Idempotency is imposed approximately, using the LNV approach\cite{Li:1993lg,Nunes:1994pi}, where the density matrix $K$ is written in terms of an \emph{auxiliary} density matrix, $L$, as $K = 3LSL - 2LSLSL$ and the elements of the auxiliary density matrix, $L^{i\alpha j\beta}$, are varied in the search for the ground state.  The initial density matrix is constructed using the Palser-Manolopoulos generalisation of the McWeeny transform\cite{Bowler:1999if,Palser:1998fd}.

The electron number is found as $N_{e}^{(K)} = 2\mathrm{Tr}[KS]$, where the superscript $^{(K)}$ is used to distinguish the calculated electron number from the desired electron number, $N_{e}$.  To optimise the energy with respect to the elements of the auxiliary density matrix, $L^{i\alpha j\beta}$, while maintaining electron number, we use a Lagrange multiplier, $\mu$, to impose the condition\cite{Hernandez:1996bf}.    As above, we have $E_{band} = 2\mathrm{Tr}\left[KH\right]$, and we define a modified band energy which includes the Lagrange multiplier:
\begin{eqnarray}
  \label{eq:6}
  \tilde{E}_{band} &=& E_{band} - \mu\left(N^{(K)}_{e} - N_{e}\right) \\
  &=& 2\mathrm{Tr}\left[KH\right] - \mu\left(2\mathrm{Tr}\left[KS\right] - N_{e}\right)
\end{eqnarray}
Naturally, when the density matrix gives the target electron number, this energy is identically equal to the band energy.  It is sometimes convenient to define $H^{\prime} = H-\mu S$ and $\tilde{E}_{band} = 2\mathrm{Tr}\left[KH^{\prime}\right] + \mu N_{e}$ to include the Lagrange multiplier in the Hamiltonian\cite{Hernandez:1996bf}, but this can obscure a contribution to the forces discussed below, and we do not follow this convention.

To find the optimum ground state, we need the derivative of this energy with respect to the auxiliary density matrix elements:
\begin{eqnarray}
  \label{eq:7}
  \frac{\partial \tilde{E}_{band}}{\partial L} &=& \frac{\partial E_{band}}{\partial L} - \mu\frac{\partial N^{(K)}_{e}}{\partial L}\nonumber\\
  \frac{\partial \tilde{E}_{band}}{\partial L} &=& \left[6\left(HLS+SLH\right)-4\left(HLSLS - SLHLS - SLSLH\right)\right] - \mu\left[12(SLS-SLSLS)\right]  
\end{eqnarray}

Here we have defined the change in electron number calculated from the $K$ matrix with respect to the elements of the auxiliary density matrix as $\partial N^{(K)}_{e}/\partial L$.  We can then set the value of the Lagrange multiplier so that, to first order, the electron number is preserved while the energy is minimised.  This is imposed by ensuring that the variation in energy is orthogonal to the variation in electron number:
\begin{eqnarray}
  \label{eq:8}
  \frac{\partial \tilde{E}_{band}}{\partial L} \cdot \frac{\partial N^{(K)}_{e}}{\partial L} &=& 0\nonumber\\
  \left(\frac{\partial E_{band}}{\partial L} - \mu\frac{\partial N^{(K)}_{e}}{\partial L}\right)\cdot \frac{\partial N^{(K)}_{e}}{\partial L} &=& 0\nonumber\\
  \Rightarrow \left(\frac{\partial E_{band}}{\partial L} \cdot \frac{\partial N^{(K)}_{e}}{\partial L}\right) / \left(\frac{\partial N^{(K)}_{e}}{\partial L} \cdot \frac{\partial N^{(K)}_{e}}{\partial L}\right) &=& \mu
\end{eqnarray}
Of course, the Lagrange multiplier, $\mu$, is the chemical potential for the electrons.

At the electronic ground state, where $\partial \tilde{E}_{band}/\partial L = 0$, we can find the contribution to the ionic forces and the stresses from the band energy (and associated Lagrange multiplier term) by considering variations in the energy with respect to the Hamiltonian and overlap matrices\cite{Miyazaki:2004ee}:

\begin{eqnarray}
  \label{eq:9}
  \tilde{E}_{band} &=& E_{band} - \mu\left(N^{(K)}_{e} - N_{e}\right) = 2\mathrm{Tr}\left[KH\right] - \mu\left(2\mathrm{Tr}\left[KS\right] - N_{e}\right)\\
  \label{eq:10}
  \delta \tilde{E}_{band} &=& \delta E_{band} - \mu \delta N^{(K)}_{e}\\
  \delta E_{band} &=& 2\mathrm{Tr}\left[\left(3L\delta SL-2L\delta SLSL-2LSL\delta SL\right)H\right] + 2\mathrm{Tr}\left[\left(3LSL-2LSLSL\right)\delta H\right]\nonumber\\
  &=& 2\mathrm{Tr}\left[K\delta H\right] -2\mathrm{Tr}\left[G \delta S\right]\\
  G &=& 2LHLSL + 2LSLHL - 3LHL\\
  \delta N^{(K)}_{e} &=& 2\mathrm{Tr}\left[\left(3L\delta SL-2L\delta SLSL-2LSL\delta SL\right)S\right] + 2\mathrm{Tr}\left[\left(3LSL-2LSLSL\right)\delta S\right]\nonumber\\
  &=& 2\mathrm{Tr}\left[\left(6LSL - 6LSLSL\right)\delta S\right]\\
\label{eq:11}
  \delta \tilde{E}_{band} &=& 2\mathrm{Tr}\left[K\delta H\right] -2\mathrm{Tr}\left[G \delta S\right] - 2\mu\mathrm{Tr}\left[\left(6LSL - 6LSLSL\right)\delta S\right] = 2\mathrm{Tr}\left[K\delta H\right] -2\mathrm{Tr}\left[\tilde{G} \delta S\right]\\
  \label{eq:12}
  \tilde{G} &=& 2LHLSL + 2LSLHL - 3LHL + \mu \left(6LSL - 6LSLSL\right)
\end{eqnarray}
The sign of the matrix $G$ is chosen for compatibility with the equivalent matrix required for force and stress calculations with exact diagonalisation:
\begin{equation}
  \label{eq:33}
  G^{i\alpha j\beta} = \sum_{n} f_{n}\epsilon_{n}c^{i\alpha\star}_{n}c^{j\beta}_{n}
\end{equation}

It is important to note that in O(N) calculations there is a contribution to the forces and stresses from the variation of the electron number with the overlap matrix which arises from the Lagrange multiplier: the final term in Eqs.~\eqref{eq:10} and \eqref{eq:11}.  This term makes an almost negligible change to the forces, typically around $1\times 10^{-4}$ Ha/a$_{0}$, and was neglected in our original formulation of forces\cite{Miyazaki:2004ee}; the effect of the term is only visible as a small variation in the energy conservation of NVE molecular dynamics runs\cite{Arita:2014ca}, and makes no difference to optimised structures.  However, the term forms a significant part of the stresses, and must be included for correct agreement between numerical and analytic stresses at small L matrix ranges.  As the $L$ matrix range increases, the contribution tends to zero, since the $L$ matrix tends towards perfect idempotency, with $LSL = L$; the small fluctuations in energy over time in NVE MD runs also reduce with increasing density matrix range, and are eliminated when this extra contribution to the force is included.

\section{Formulating stress with localised orbitals and linear scaling}

We have separated the contributions to the stress into different types, in a similar way to the presentation of forces\cite{Miyazaki:2004ee}.  Hellmann-Feynman stresses arise from explicit dependence of the potentials on the atom positions via the strain, while Pulay stresses come from the change in the basis when atoms move (since the basis is atom-centred).  There are then further miscellaneous stresses which are important but do not fit neatly into any category.  We note that the calculation of the stress requires negligible time, as it is included with the force evaluation (this is standard for local orbital codes, in contrast to plane wave basis sets).

\subsection{Hellmann-Feynman stress}
\label{sec:hellm-feynm-stress}

The Hellmann-Feynman stress is straightforward, and equivalent to other codes, coming from the pseudopotentials; there is an extra contribution from the screened Hartree potential in the neutral atom formulation.
\begin{equation}
    \label{eq:13}
    \sigma^{HF}_{\lambda\mu} = \sum_{i}\int d\mathbf{r} n(\mathbf{r})\frac{\partial V^{NA}_{i}(\mathbf{r})}{\partial R_{i\lambda}} R_{i\mu} - 2\sum_{i}\sum_{ab}K_{ab}\langle\phi_{a}\mid \left(\nabla_{i\lambda}\chi_{b}\right)R_{i\mu}\rangle A_{i}\langle \chi_{i}\mid\phi_{b}\rangle - \sum_{i}\int d\mathbf{r} \delta V_{Ha}(\mathbf{r})\frac{\partial n^{PAD}_{i}}{\partial R_{i\lambda}}R_{i\mu}
  \end{equation}
To understand why we classify the final term as a Hellmann-Feynman term, we note that the screened Hartree energy is written as $E_{\delta Ha} = \frac{1}{2}\int d\mathbf{r} \delta V_{Ha} \left(n(\mathbf{r}) - \sum_{i} n_{i}^{PAD}(\mathbf{r})\right)$, so there is a contribution to the force and stress from the change of $n_{i}^{PAD}(\mathbf{r})$ with atomic position.  The equivalence to a Hellmann-Feynman force comes because the integral can be recast as $\delta n(\mathbf{r})$ interacting with $V_{Ha}^{PAD}(\mathbf{r})$, the potential due to the atomic density.
  
\subsection{Pulay stresses}
\label{sec:pulay-stresses}

Following the forces\cite{Miyazaki:2004ee}, we find that there are two contributions to the Pulay stress, coming from changes in the Hamiltonian and overlap matrices; we term these $\phi$-Pulay and S-Pulay stresses, respectively.  In Eq.~\eqref{eq:15} below, the first three terms form the $\phi$-Pulay stress (contributions from the local potentials, i.e. neutral atom, Hartree and XC potentials, in the first term; the kinetic energy in the second term; and the non-local potential in the third term) while the fourth contribution forms the S-Pulay stress.
  \begin{eqnarray}
    \label{eq:15}
    \sigma^{Pulay}_{\lambda\mu} &=& -4\sum_{i}\sum_{ab}K_{ab}\langle\phi_{a}\mid \hat{V}^{NA} + \delta\hat{V}^{Ha} + \hat{V}^{XC}\mid \left(\nabla_{i\lambda}\phi_{b}\right)R_{i\mu}\rangle  \nonumber\\
                                &+&\frac{1}{2}\sum_{i}\sum_{ab}K_{ab}\frac{\partial T_{ab}}{\partial R_{i\lambda}} R_{i\mu}-2\sum_{i}\sum_{abk}K_{ab}\langle\left(\nabla_{i\lambda}\phi_{a}\right)\mid \chi_{k}\rangle R_{i\mu} A_{k}\langle \chi_{k}\mid\phi_{b}\rangle\nonumber\\
    &-&4\sum_{i}\sum_{ab}\tilde{G}_{ab}\langle\phi_{a}\mid \left(\nabla_{i\lambda}\phi_{b}\right)\rangle R_{i\mu}
  \end{eqnarray}
  It is important to note that, for linear scaling, the matrix $\tilde{G}_{ab}$ is the same matrix found in Eq.~\eqref{eq:12}, including the effect of the variation of the electron number, while for exact diagonalisation it is simply $G_{ab}$.

\subsection{Miscellaneous Stresses}
\label{sec:misc-stress}

The remaining stresses do not fit easily into the two previous categories, and are given here in no particular order.  The ion-ion stress within the neutral atom potential is written as follows:
\begin{eqnarray}
  \label{eq:16}
  \sigma^{SCC}_{\lambda\mu} &=& \sum_{i}\frac{1}{2}\sum_{j}\left[\frac{Z_{i}Z_{j}}{\mid\mathbf{R}_{i} - \mathbf{R}_{j}\mid^{3}}\left(R_{i\lambda} - R_{j\lambda}\right)\delta_{\lambda\mu} - \int d\mathbf{r} V^{Ha,PAD}_{j}(\mathbf{r})\frac{\partial n^{PAD}_{i}}{\partial R_{i\lambda}}\right]\left(R_{i\mu} - R_{j\mu}\right)
\end{eqnarray}
  
Various terms in the total energy are calculated using a numerical integration on a real-space grid (the screened Hartree, exchange-correlation and neutral atom energies).  There is a stress associated with the change of the Jacobian (i.e. the grid point volume):
\begin{equation}
  \label{eq:17}
  \sigma^{Jacobian}_{\lambda\mu} = \delta_{\lambda\mu}\left(E^{\delta Ha} + E^{XC} + E^{NA}\right)
\end{equation}
Similarly, since the Hartree potential is found using fast Fourier transforms, there is a stress which arises from changes of the reciprocal lattice vectors:
  \begin{equation}
    \label{eq:19}
    \sigma^{Ha}_{\lambda\mu} = \int d\mathbf{r} \delta n(\mathbf{r}) FT\left[\frac{2G_{\lambda}G_{\mu}}{G^{4}}\tilde{\delta n}\right]
  \end{equation}
where $\tilde{\delta n}$ indicates the Fourier transform of $\delta n$, and $FT[]$ indicates the inverse Fourier transform.  

For a GGA exchange-correlation functional, which depends on the gradient of the electron density, there is an extra contribution to the stress which can be written as\cite{Dal-Corso:1994os,Balbas:2001it}:
  \begin{equation}
    \label{eq:20}
    \sigma^{XC, GGA}_{\lambda\mu} = \int d\mathbf{r}\frac{\partial f}{\partial \nabla_{\lambda}n(\mathbf{r})}\nabla_{\mu}n(\mathbf{r})
  \end{equation}

When using non-linear partial core corrections (PCC), there is a stress associated with the change of the core densities as atoms move, which can be written as:
\begin{equation}
  \label{eq:18}
  \sigma^{PCC}_{\lambda\mu} = \sum_{i} \int d\mathbf{r} \nabla_{i\lambda} n^{C}_{i}(\mathbf{r}) V_{XC}\left[n(\mathbf{r}) + n^{C}(\mathbf{r})\right] R_{i\mu}
\end{equation}
It might be argued that this is similar to a Hellmann-Feynman stress, but the distinction is not important.

Finally, when running without self-consistency, there are stresses equivalent to the forces we have described elsewhere\cite{Torralba:2009nr}:
\begin{eqnarray}
  \label{eq:21}
  \sigma^{Non-SCF}_{\lambda\mu} &=& -\int d\mathbf{r}\Delta V^{Ha}(\mathbf{r})\sum_{i}\frac{\partial }{R_{i\lambda}}n^{PAD}_{i}R_{i\mu}\nonumber\\
                                &-& \int d\mathbf{r}\Delta n(\mathbf{r})\mu_{XC}^{\prime}\left[n^{PAD} + n^{C}\right]\sum_{i}\frac{\partial }{R_{i\lambda}}\left(n^{PAD}_{i} + n^{C}_{i}\right)R_{i\mu}
\end{eqnarray}
There is also an extra Jacobian stress for the exchange and correlation:
\begin{equation}
  \label{eq:14}
  \sigma^{XC Jacobian, Non-SCF} = \int d\mathbf{r}\left(n^{out}-n^{PAD}\right)V_{XC}[n^{PAD} + n^{C}]
\end{equation}
The Hartree stress arising from the change in reciprocal lattice vectors is identically zero for non-self-consistent calculations.

\section{Tests and results}

In this section we present comprehensive tests of the convergence of
pressure, force and total energy with density matrix range for a
variety of materials with differing properties, and then test
molecular dynamics in the NPT ensemble.  In all cases we compare
results found with linear scaling to those found with converged
exact diagonalisation.  The convergence of total energy with respect
to density matrix range is related to the
gap of the material\cite{Prodan:2005ql};
while there are no formal results on the convergence of
  forces or stresses with density matrix range, energy differences 
  converge faster than total energies, and most linear
scaling calculations rely on the fast convergence of energy
\emph{differences}, even when total energies are not converged.  It
is important to note that the same is true
for many parameters in standard DFT calculations, e.g. k-point
convergence, or plane wave cutoff in plane wave codes.
For simplicity, we use a simple single zeta
basis set of pseudo-atomic orbitals\cite{Bowler:2019fv} for all
calculations; 
the only significant effect that a larger basis set will
  have on the results will be if the gap is changed by the basis.  Any basis
  set superposition errors (BSSE) will have a negligible effect on stress
  calculations in condensed phases.

All tests used Conquest v1.5\cite{Conquest:2026fb} which is publicly available from GitHub, and includes the stress implementation described in this paper.

\subsection{Convergence Tests}
\label{sec:convergence-tests}

We consider six materials which provide a variety of types of bonding
and electronic structure.  The elemental semiconductors
carbon, silicon and germanium are all tetrahedrally bonded, and have
varying gaps (the values with the single zeta basis are not accurate,
but nevertheless span an appropriate range: 5.9eV for carbon; 2.1eV for
silicon; and 0.0eV for germanium at the $\Gamma$ point), covering behaviours
from insulating,
through semiconducting to metallic.  We also include ionic materials,
all of which have large gaps: MgO and LiCl, which are both cubic,
with gaps of 5.3eV and 7.0eV; and
AlN, which is hexagonal (wurtzite), and polar along the $c$-axis, with a
gap of 3.4eV.  This
variety of materials covers differences in bonding and electronic
structure, and provides a stringent test of the approach.  For MgO and LiCl
we use the local density approximation with Troullier-Martins
pseudopotentials\cite{Troullier:1991ai}, while for the other
materials we use PBE\cite{Perdew:1996sj} with Hamann
pseudopotentials\cite{Hamann:2013kq} from the PseudoDojo
database\cite{Setten:2018xv}.

In our comparisons, we use the same simulation cells, with all parameters
identical except for the ground state search.  In all cases, we used a
perfect crystal but with one atom
slightly displaced along the c-axis of the cell to give non-zero forces,
with a lattice constant
reduced by 1\% from the relaxed value (for exact diagonalisation).
We use a real space integration grid with a spacing of 0.2a$_{0}$,
which gives good convergence.  All diagonalisation calculations used
a Monkhorst-Pack grid which fully converged the total energy.

\begin{figure}
    \centering
    (a)
    \includegraphics[width=0.45\columnwidth]{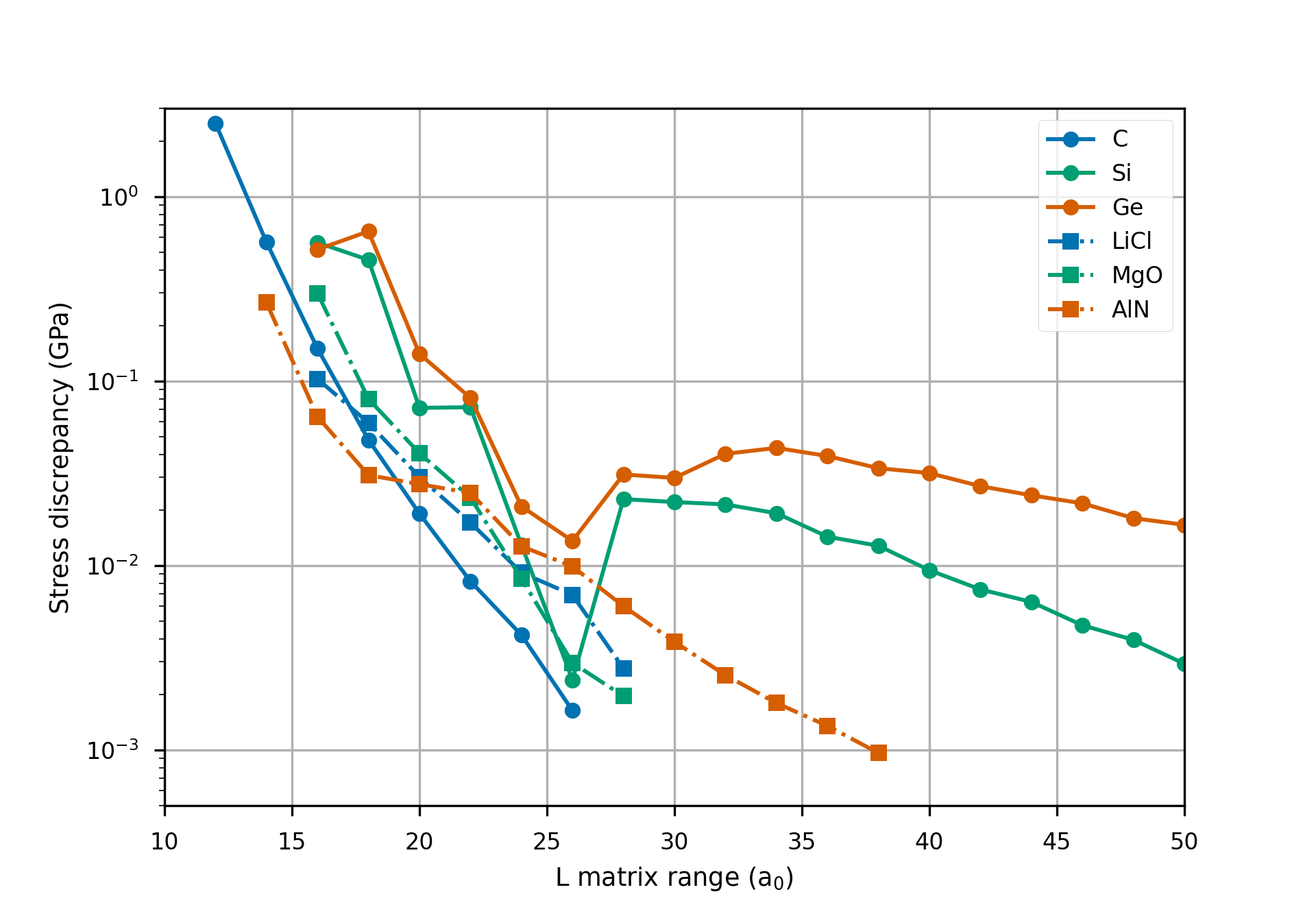}
    (b)
    \includegraphics[width=0.45\columnwidth]{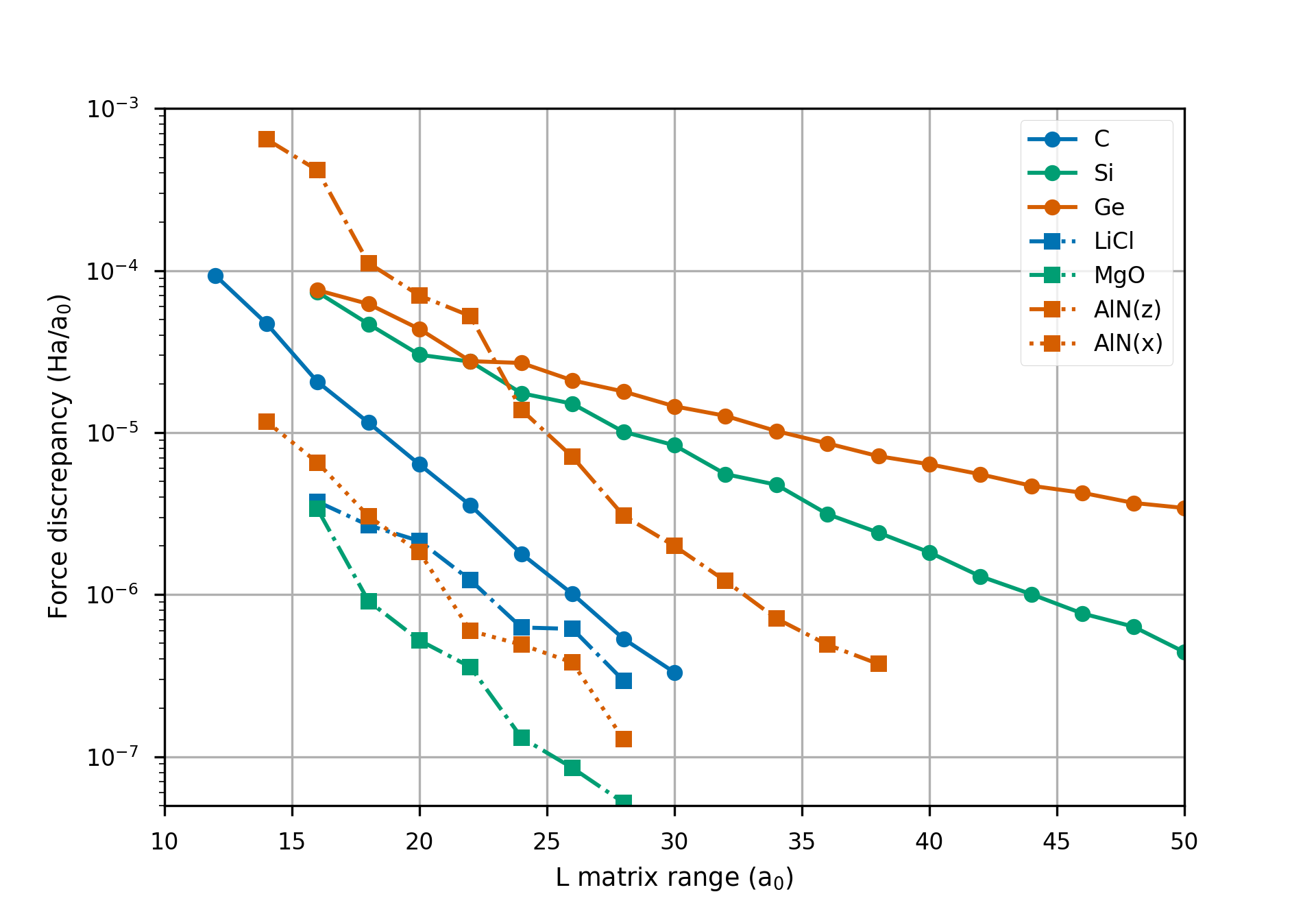}\\
    (c)
    \includegraphics[width=0.45\columnwidth]{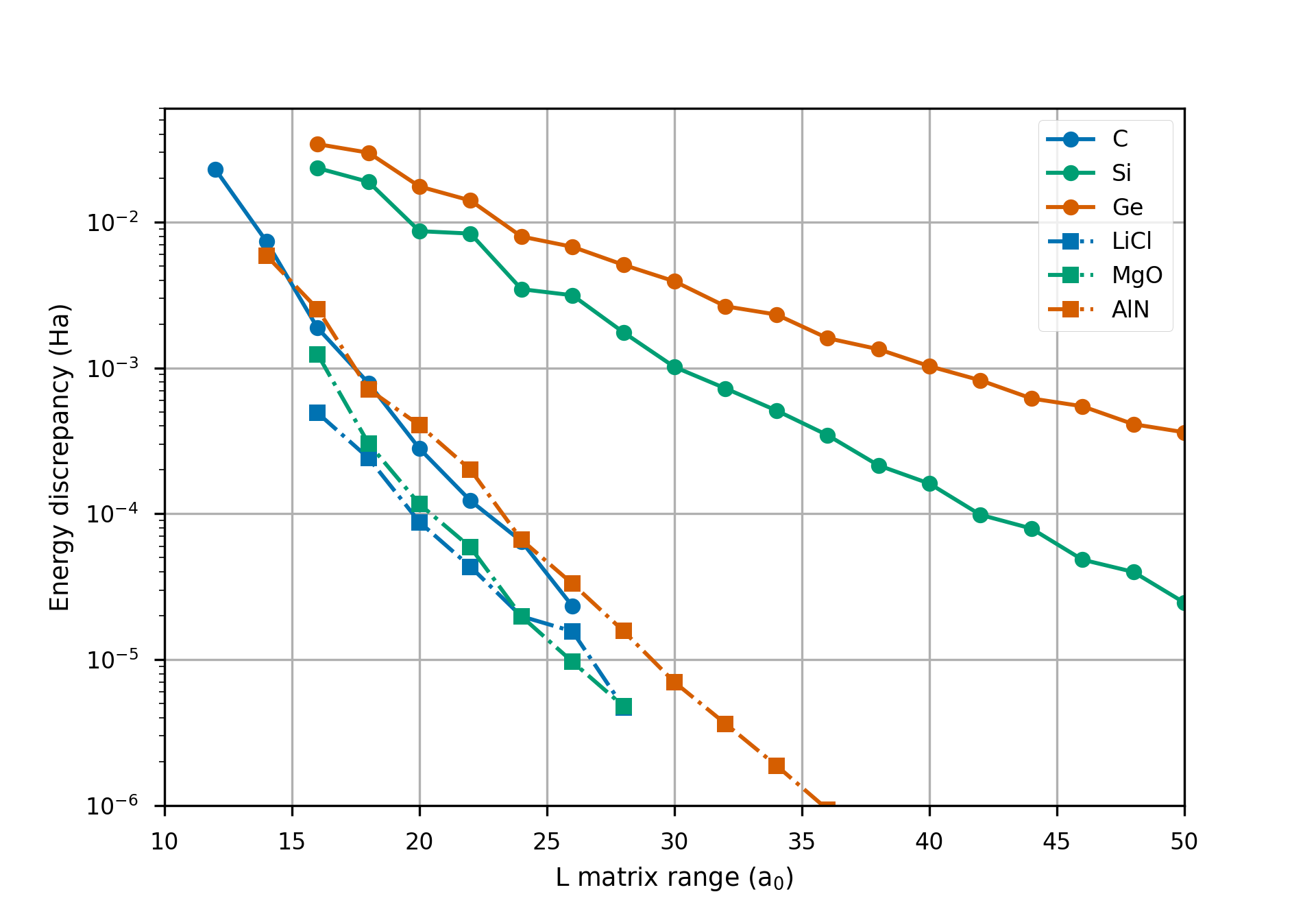}
    (d)
    \includegraphics[width=0.45\columnwidth]{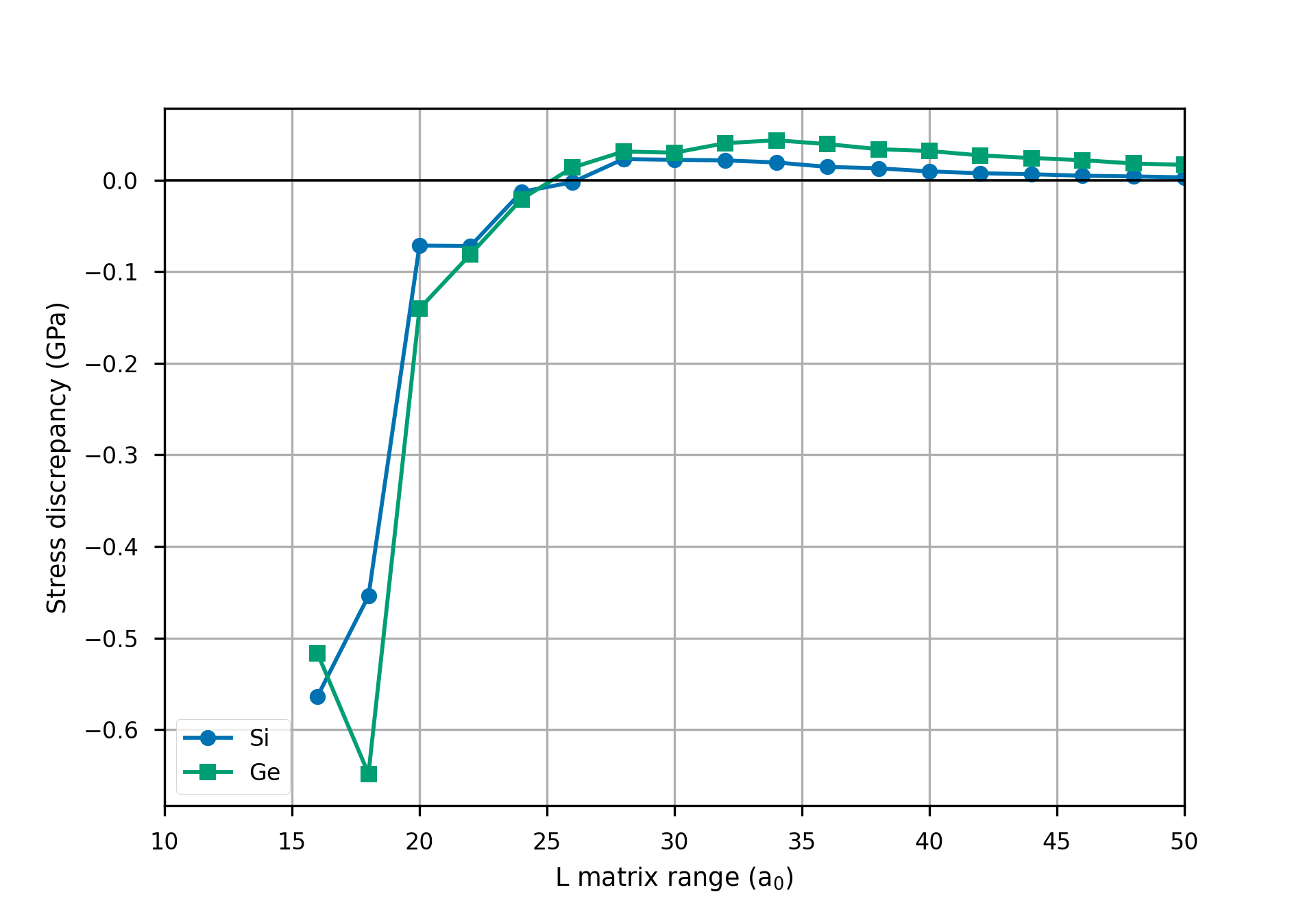}
    \caption{Absolute value of the difference between linear scaling and fully converged exact diagonalisation calculations as a function of density matrix cutoff (in a$_{0}$) for: (a) stress (in GPa); (b) force (in Ha/a$_{0}$); and (c) total energy (in Ha). (d) stress difference for Si and Ge, as in (a), on a linear scale to illustrate the cause of the dip in the stress difference curve.}
    \label{fig:PressErrorBohr}
  \end{figure}

We consider three
quantities: the total energy; the force; and the stress along the c-axis.  In all
practical circumstances we are interested in energy \emph{differences} rather
than total energies, so the stress and force graphs are much more
important.  The results are shown in Fig.~\ref{fig:PressErrorBohr},
showing the stress in (a), the force in (b) and the total energy in
(c).  The unit cells of all materials apart from AlN are cubic; we have plotted the
stress along the $c$ axis in AlN, which is the direction of the
displacement of the ion, though the convergence is similar in all directions.

Considering first the stress, we see that the discrepancy is rather
small, being below 1GPa for all materials at all density matrix ranges
(except carbon with the exceptionally small density matrix range of
12\,a$_{0}$) and below 0.1GPa for very reasonable density matrix ranges:
18\,a$_{0}$ for the four insulators (C, LiCl, MgO and AlN), 20\,a$_{0}$ for
silicon, and 22\,a$_{0}$ for germanium.  The dip in the curve seen around
25\,a$_{0}$ for Si and Ge is due to the non-monotonic convergence of the
stress with density matrix range, as shown in
Fig.~\ref{fig:PressErrorBohr}(d); there is no fundamental reason why
the stress should converge monotonically.

The force convergence is even better than the stress, with the discrepancies below 1$\times$10$^{-4}$Ha/a$_{0}$ ($\simeq5\times 10^{-3}$ eV/\AA) for all density matrix ranges and all materials (except AlN along a direction in the basal plane where this level requires a range of 18a$_{0}$).  This indicates that structural optimisation for all elements considered here will be reliable regardless of the restriction placed on the density matrix.  The final curve, showing convergence of total energy, displays the expected trends of exponential convergence, with larger gaps converging faster.  Absolute energies are much less important than energy differences, which converge much faster (as seen in the stresses and forces). We note that comparison of the analytic stresses presented here with numerical stresses found from the energy change resulting from a small volume change are in excellent agreement.

As a further test of the stress calculation, we evaluate the bulk modulus for all materials by fitting to seven point curves in both the energy (as is standard for DFT calculations) and the pressure on the simulation cell (which tests the stress explicitly).  For the linear scaling calculations, we use the same density matrix range identified above where the stress error is less than 0.1GPa (18$a_{0}$ for C, LiCl, MgO and AlN; 20$a_{0}$ for Si and 22$a_{0}$ for Ge).  We found that rather than using a fixed density matrix range for each point on the binding energy curve, better convergence is obtained by scaling the density matrix range with simulation cell size\footnote{This ensures that the number of non-zero density matrix elements is constant across all calculations, in close analogy to similar calculations performed with a plane wave basis set, which generally use a constant number of plane waves.}.  The results are given in Table~\ref{tab:B0}, and show excellent agreement.

\begin{table}[h]
  \centering
  \begin{tabular}{lcccccc}
                      &     C &   Si &   Ge & LiCl &   MgO &   AlN \\
    O(N) B$^{(E)}_{0}$ & 391.8 & 72.5 & 46.0 & 39.8 & 187.3 & 182.1 \\
    Diag B$^{(E)}_{0}$ & 391.8 & 71.3 & 46.9 & 39.6 & 185.5 & 181.7 \\
    O(N) B$^{(P)}_{0}$ & 395.6 & 71.9 & 46.2 & 39.3 & 185.6 & 180.9 \\
    Diag B$^{(P)}_{0}$ & 395.5 & 70.5 & 47.9 & 39.1 & 185.4 & 180.4
  \end{tabular}
  \caption{Bulk modulus in GPa calculated by fitting Birch-Murnaghan curves to energy (B$^{(E)}_{0}$) and pressure (B$^{(P)}_{0}$) for the six materials, for linear scaling ($R_{L} = 18a_{0}$ for C, LiCl, MgO, AlN, $R_{L} = 20a_{0}$ for Si and $R_{L} = 22a_{0}$ for Ge) and exact diagonalisation.}
  \label{tab:B0}
\end{table}
\subsection{Molecular Dynamics Tests}
\label{sec:molec-dynam-tests}

We now turn to a practical application of the calculated stress: isobaric-isothermal molecular dynamics.  We have already demonstrated NVE molecular dynamics with linear scaling DFT on systems up to 32,678 atoms\cite{Arita:2014ca}, using the extended Lagrangian Born-Oppenheimer (XL-BOMD) approach\cite{Niklasson:2008dz}.  In that work, we showed that the total energy was well conserved, with small fluctuations (no more than 1 meV/atom even for a small density matrix range of 14 a$_{0}$) which reduced with increasing density matrix range.  We have also demonstrated dynamics under the NVT ensemble\cite{Hirakawa:2017pt}, again showing stable dynamics with good conservation.  Here, we show that stable linear scaling molecular dynamics is possible for the NPT ensemble, and characterise the behaviour with density matrix range.  We choose a temperature of 300K and a pressure of 10GPa for a sample of bulk silicon.  We are particularly interested in the effects of the density matrix range on fidelity of the dynamics: both the stability and conservation of the simulations, and how closely the full diagonalisation dynamics can be followed.

\begin{figure}
  \centering
    (a)
  \includegraphics[width=0.3\columnwidth]{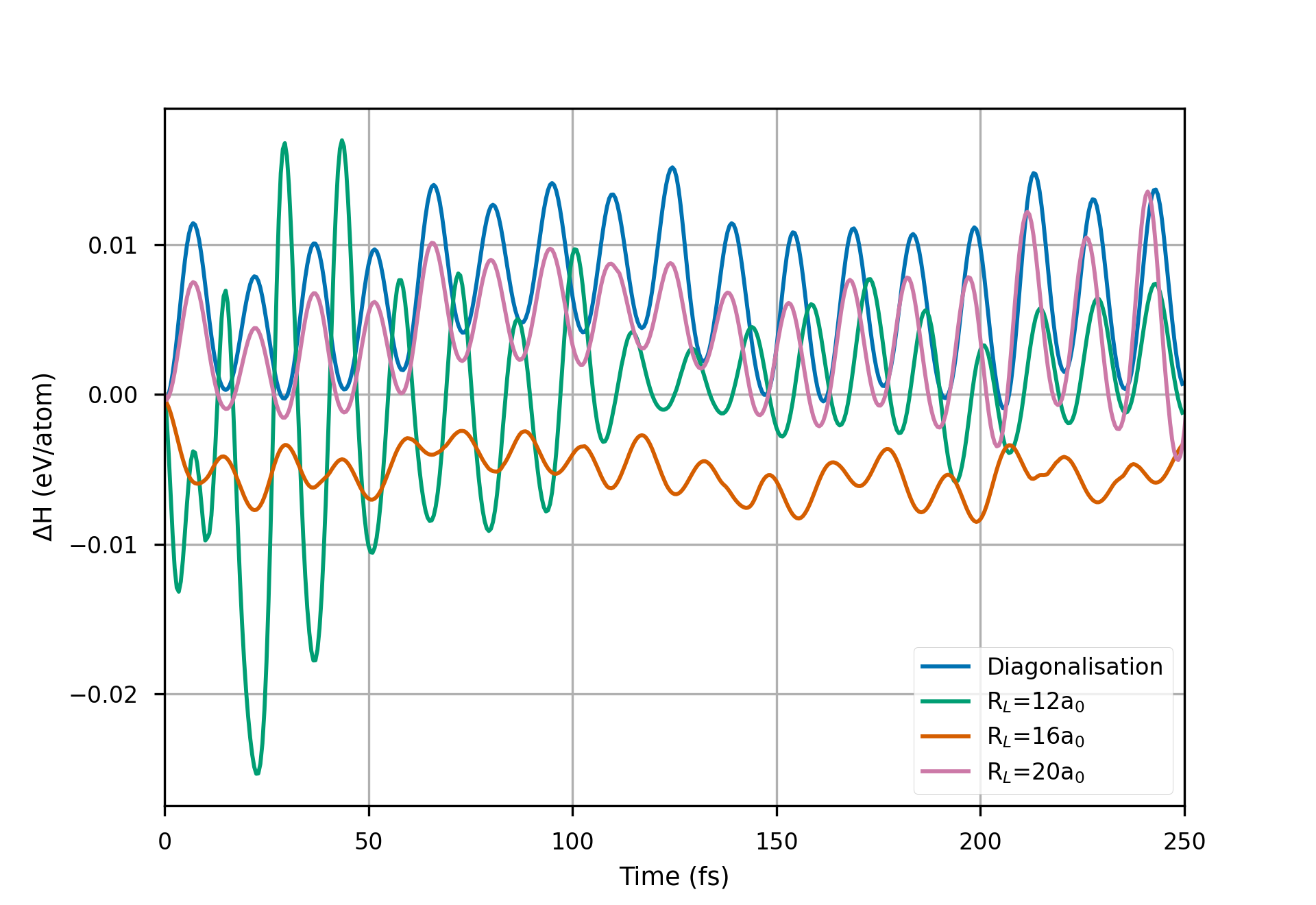}
    (b)
  \includegraphics[width=0.3\columnwidth]{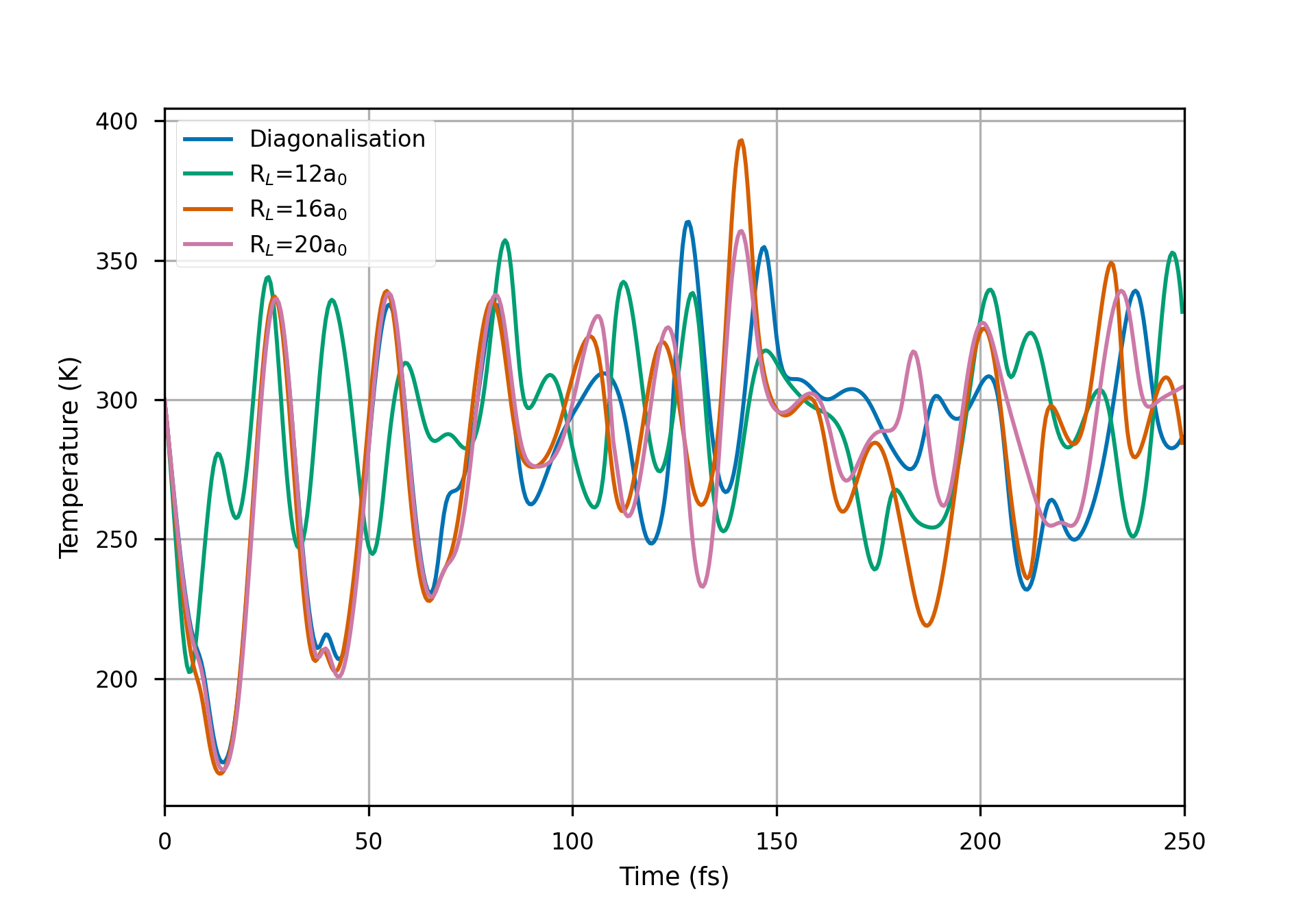}
    (c)
  \includegraphics[width=0.3\columnwidth]{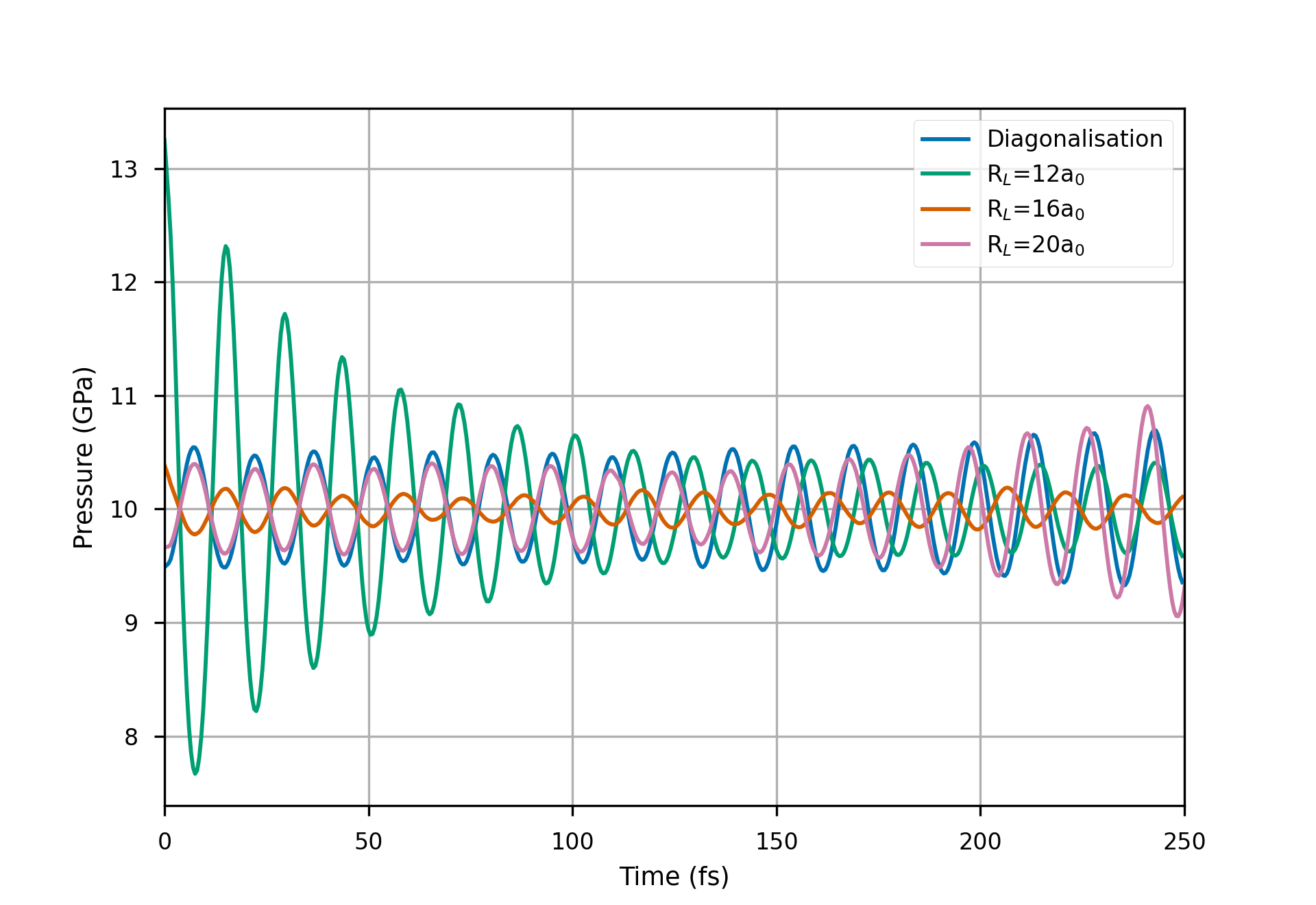}
  \caption{Characterisation of the effect of density matrix range on NPT molecular dynamics.  (a) Deviation of conserved quantity; (b) temperature and (c) pressure variation.  In all cases results for full diagonalisation are compared to three density matrix ranges (12, 16 and 20a$_{0}$).}
  \label{fig:CompareMD}  
\end{figure}

In Fig.~\ref{fig:CompareMD} we consider the case where we used identical simulation settings for all tests, including the same starting seed for the pseudo-random number generator and starting simulation cell size.  We note that the initial pressures in these cases differ with density matrix range (as seen in Fig.~\ref{fig:PressErrorBohr}), with a difference to exact diagonalisation of less than 0.1GPa at $R_{L}=20a_{0}$.  As the density matrix range increases, the linear scaling dynamics converge towards the exact diagonalisation dynamics; in all cases, we see that the conserved quantity does not drift, and the pressure and temperature are maintained at their set point.  

\begin{figure}
  \centering
    (a)
  \includegraphics[width=0.3\columnwidth]{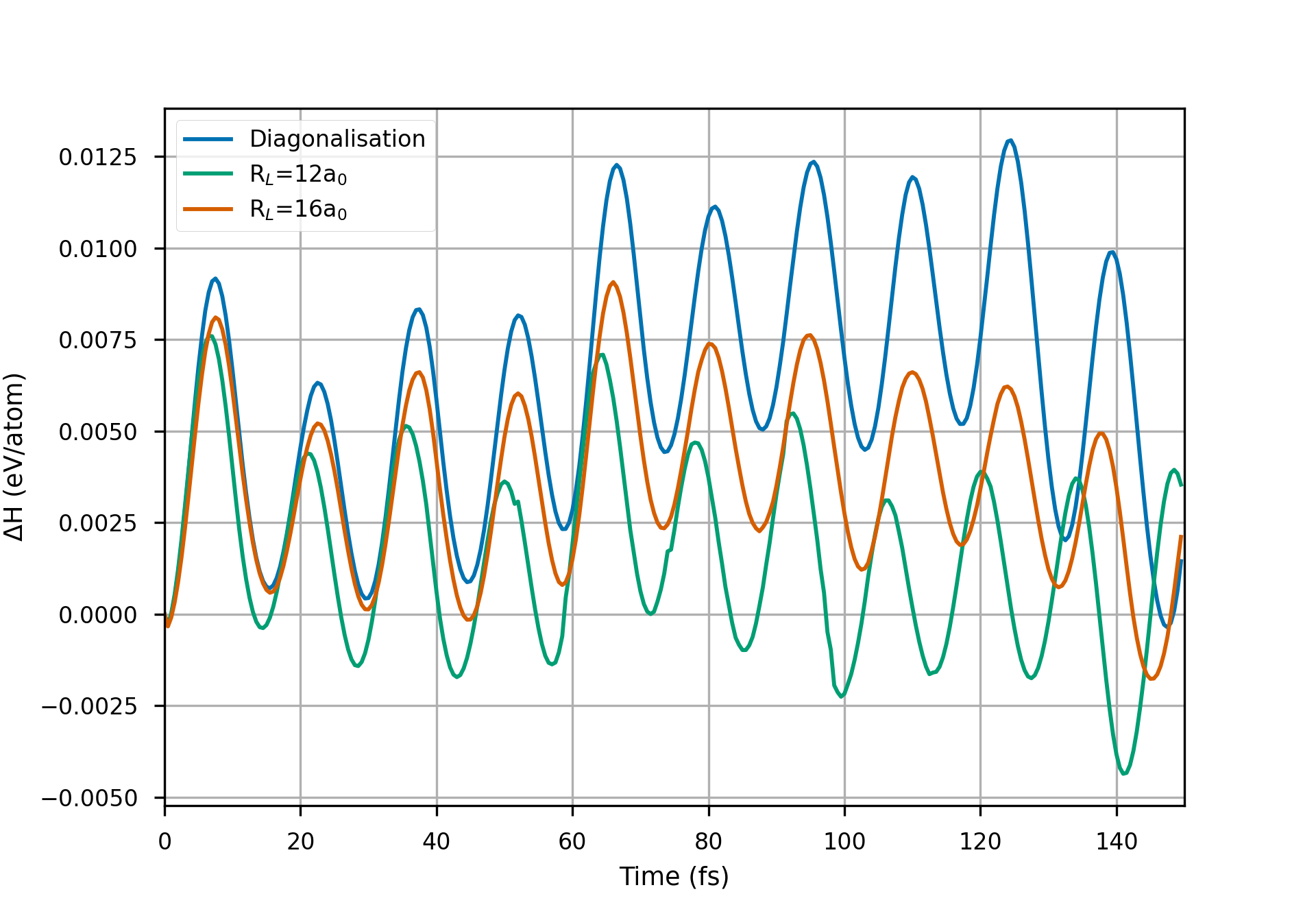}
    (b)
  \includegraphics[width=0.3\columnwidth]{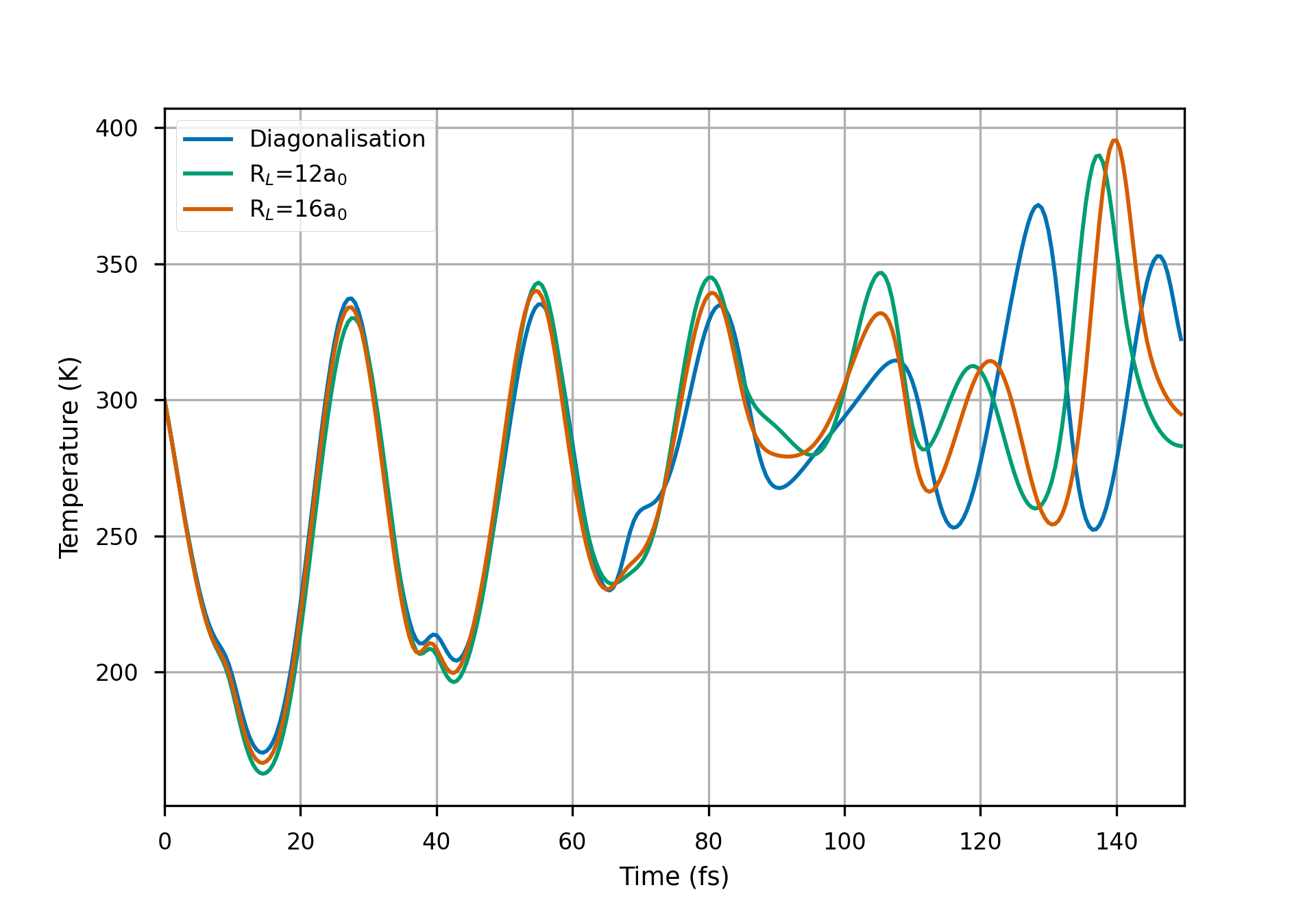}
    (c)
  \includegraphics[width=0.3\columnwidth]{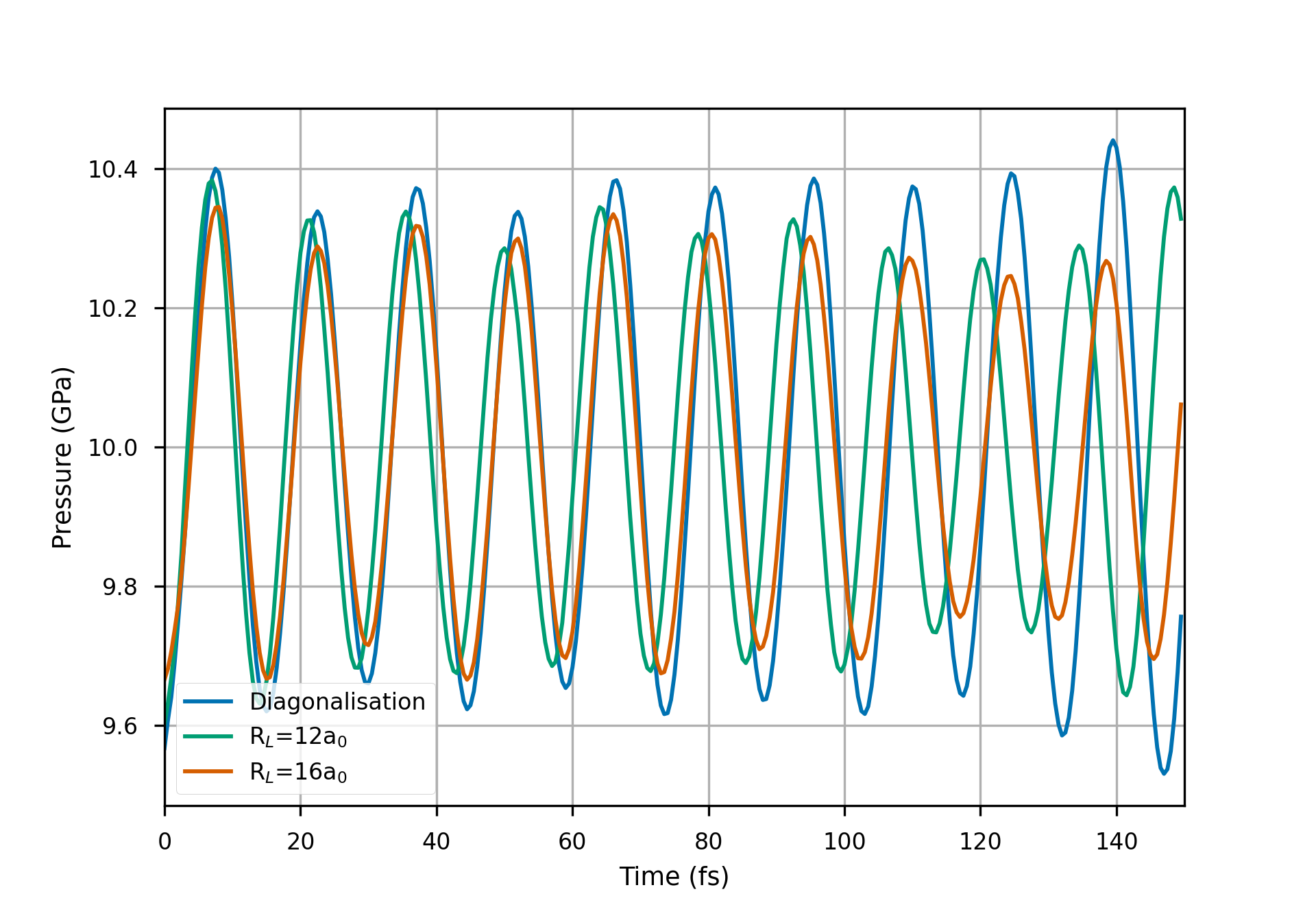}
  \caption{Comparison of NPT molecular dynamics starting from the same initial pressure for diagonalisation and linear scaling calculations.  (a) Deviation of conserved quantity; (b) temperature and (c) pressure variation.  In all cases, results for full diagonalisation are compared to two density matrix ranges (12 and 16a$_{0}$).}
  \label{fig:CompareSamePress}
\end{figure}

As would be expected, the smaller density matrix ranges show more differences in the detailed dynamics to the exact diagonalisation results, while the $R_{L}$=20a$_{0}$ results follow the exact diagonalisation results closely.  In particular, the pressure fluctuations are out of phase with the exact diagonalisation results, and the conserved quantity follows a somewhat different path (though with the same underlying frequency of oscillation).  The key question to understand in terms of fidelity of dynamics is whether this is purely due to the difference in initial pressure, or also due to differences in forces; from the results in Fig.~\ref{fig:PressErrorBohr}(b) we would expect it to be almost entirely from pressure.  We investigated dynamics for short density matrix ranges ($R_{L}$=12 and 16a$_{0}$) with simulation cell sizes adjusted to give the same initial pressure as in the exact diagonalisation simulation (to within 0.1GPa).  The results are shown in Fig.~\ref{fig:CompareSamePress}.

We see that for the first hundred femtoseconds of dynamics the short density matrix range simulations follow the exact diagonalisation trajectories extremely closely.  The increase in difference over time is an inevitable result of slight variations in dynamics.  The result of these simulations is extremely encouraging for large scale, long-time DFT dynamics: we see that high quality simulations can be achieved using very modest density matrix ranges.  Comparing diagonalisation and O(N) calculations\footnote{To simplify this comparison, we exclude the thermostat energy so that any difference is solely due to the electronic structure solver.} we find a drift of less than 0.001eV/ps/atom for simulations spanning multiple picoseconds.

  \section{Conclusions}

  We have shown how stress can be calculated in the context of the local orbital DFT code \textsc{Conquest}, both in terms of exact diagonalisation and linear scaling operation.  Tests of the convergence of the stress calculated with linear scaling to the exact diagonalisation result show that relatively modest density matrix ranges are required for excellent agreement, with all materials converged to within 0.1GPa with a range of 20a$_{0}$.  Tests of NPT dynamics show no drift in the conserved quantity, and stable dynamics over meaningful timescales.  Moreover, adjusting the starting simulation cell size so that the initial pressures of both linear scaling and exact diagonalisation are the same allows us to reproduce the exact dynamics over 100fs for very small ranges, as low as 12a$_{0}$ for silicon.

  Our results demonstrate that the full range of molecular dynamics ensembles can be used with linear scaling DFT methods with no loss of accuracy in the dynamics of the system, even with modest density matrix ranges.  This opens the prospect of very large scale dynamics calculations with full DFT accuracy.  The remaining questions for linear scaling MD simulations concern basis set size and overall efficiency.  Since the support functions in \textsc{Conquest} are non-orthogonal, we require the inverse of the overlap matrix for efficient calculations\cite{Nunes:1994pi,Bowler:2000qc}, but with linear scaling time.  At present, we find an approximate, short range inverse matrix using Hotelling's method, though this approach performs poorly for larger pseudo-atomic orbital basis sets as the overlap matrix becomes increasingly ill-conditioned.  The lack of an inverse overlap matrix potentially restricts the basis size that can be used, though the use of the alternative blip basis sets\cite{Hernandez:1996bf,Bowler:2002pt} and the on-site support function approach outlined recently\cite{Nakata:2020dn} offer ways forward here.  The question of efficiency is key to long timescale MD runs.  We already use the extended Lagrangian approach to molecular dynamics\cite{Niklasson:2008dz} to ensure energy conservation, though this can require more iterations to find the ground state than simple reuse of the density matrix; extensions of the extended Lagrangian approach that require no self-consistency cycles\cite{Souvatzis:2014uq} may help with efficiency.  Once these questions have been addressed, the route to very large scale, long DFT molecular dynamics runs is open.

  \acknowledgements{This research was partly carried out using the high performance computing cluster in the London Centre for Nanotechnology.  The authors acknowledge the use of the UCL Kathleen High Performance Computing Facility (Kathleen\@UCL), and associated support services, in the completion of this work.  We are grateful for computational support from the UK Materials and Molecular Modelling Hub, which is partially funded by EPSRC (EP/T022213/1, EP/W032260/1 and EP/P020194/1), for which access was obtained via the UKCP consortium and funded by EPSRC grant ref EP/X035891/1.}
  
  \bibliography{onstress}

\end{document}